\documentclass[3p]{elsarticle}

\usepackage{amsmath}
\usepackage{amsthm}
\usepackage{amssymb}
\usepackage{graphicx}
\usepackage{esint}
\usepackage{color}

\usepackage{lineno}
\PassOptionsToPackage{hyphens}{url}
\usepackage[pdfencoding=auto,psdextra]{hyperref}
\usepackage{bookmark}

\biboptions{numbers,sort&compress}

\modulolinenumbers[1]

\journal{Biosystems}

\begin{document}

\begin{frontmatter}

\title{Inner privacy of conscious experiences and quantum information}

\author[address1]{Danko D. Georgiev\corref{mycorrespondingauthor}}
\ead{danko.georgiev@mail.bg}
\cortext[mycorrespondingauthor]{Corresponding author}

\address[address1]{Institute for Advanced Study, Varna, Bulgaria}

\begin{abstract}
The human mind is constituted by inner, subjective, private, first-person conscious experiences that cannot be measured with physical devices or observed from an external, objective, public, third-person perspective. The qualitative, phenomenal nature of conscious experiences also cannot be communicated to others in the form of a message composed of classical bits of information. Because in a classical world everything physical is observable and communicable, it is a daunting task to explain how an empirically unobservable, incommunicable consciousness could have any physical substrates such as neurons composed of biochemical molecules, water, and electrolytes. The challenges encountered by classical physics are exemplified by a number of thought experiments including the inverted qualia argument, the private language argument, the beetle in the box argument and the knowledge argument. These thought experiments, however, do not imply that our consciousness is nonphysical and our introspective conscious testimonies are untrustworthy. The principles of classical physics have been superseded by modern quantum physics, which contains two fundamentally different kinds of physical objects: unobservable quantum state vectors, which define what physically exists, and quantum operators (observables), which define what can physically be observed. Identifying consciousness with the unobservable quantum information contained by quantum physical brain states allows for application of quantum information theorems to resolve possible paradoxes created by the inner privacy of conscious experiences, and explains how the observable brain is constructed by accessible bits of classical information that are bound by Holevo's theorem and extracted from the physically existing quantum brain upon measurement with physical devices.
\end{abstract}


\date{October 3, 2019}

\end{frontmatter}


\section{Introduction}

We are \emph{sentient} beings living inside a physical world \cite{Kirk1974,Nagel1974,Noren1973}. We are curious to know who we are, what we are, why we are, and how we relate to the surrounding world. To satisfy our curiosity by finding the answers of these questions, we can only rely on our introspective testimony of how the world appears to us as experienced through our five senses (taste, sight, touch, smell, and hearing) and what it feels like to be us. From the memorized past experiences, we construct theories of the physical world, and if these theories capture correctly certain aspects of the physical reality, we are able to predict new facts or explain new phenomena, which lie beyond the original data used to construct those theories \cite{Duhem1962,Popper2002,Rosenkrantz1977}. Since we all live in the same world, we can learn from each other and through rational reasoning, it should be possible to converge onto a single scientific theory that explains all available facts about ourselves and the physical world \cite{Georgiev2017}. While such a united scientific description of physical reality is the ultimate goal of scientific inquiry \cite{Neurath1983}, the inner privacy of conscious experiences has often been used as an argument for the ``nonphysical'' nature of consciousness \cite{Chalmers1995a,Chalmers1996,Jackson1982,Jackson1986,Robinson1976,Sprigge1994,Zhao2012} and purported failure of physics to address questions related to our sentience and mentality \cite{Kim1998,Nagel1965,Nagel1974}.

Here, we will analyze a number of theoretical arguments related to the inner privacy of consciousness, examine their soundness in the frameworks of classical or quantum physics, and show how a quantum information-theoretic approach to consciousness is able to produce a physical theory of mind that is consistent with both the fundamental physical laws and our introspective testimonies of what our conscious minds are.

\section{The inner privacy of consciousness}

The essence of \emph{consciousness} is the act of \emph{experience} \cite{Chalmers1995a,Chalmers1995b,Nagel1974}. Conscious experiences are private and accessible through introspection from a first-person, subjective, phenomenal perspective, but remain unobservable from a third-person, objective perspective \cite{Nagel1974,Nagel1987}. As a consequence, there is no objective scientific way to determine if any other person, animal or object is conscious or not, because we do not have direct access to someone else's experiences through observation. For example, we may observe someone else's brain but the process of observation alone does not make us experience what that brain is experiencing. Thus, there are some things that exist in the universe, such as one's own experiences, but which are fundamentally \emph{unobservable}. Furthermore, the phenomenal nature (qualia) of individual conscious experiences is \emph{incommunicable}. In other words, we may define the subject whose experiences we are interested in or the situation under which certain experiences are elicited; however, we are unable to describe in words what is it like to have those experiences. Therefore, we are unable to communicate what is it like to have any of our experiences to others, and those others are also unable to communicate what is it like to have their experiences to us.

The \emph{inner privacy} of consciousness intertwines the two closely related, yet different, physical concepts: \emph{unobservability} (the property of not being observable or measurable) and \emph{incommunicability} (the property of not being able to be communicated). In consciousness research, the relationship between these two concepts is the following: if conscious experiences were observable, then we would have been able to communicate the phenomenal nature of conscious experiences by doing nothing besides letting others observe us; alternatively, if conscious experiences were communicable, we would have been able to restore missing senses through words alone. Evidence in favor of unobservable and incommunicable conscious experiences is abundant and obtainable whenever you are in the company of other conscious beings including family, friends, or colleagues at work. Further insights into the meaning of unobservability and incommunicability could be attained by considering several thought experiments, which are richly informative and easy to imagine.

\subsection{Unobservability of conscious experiences}

\paragraph{What is it like to hear music}
Imagine a person who was born deaf and never knew what is it like to hear sounds. Equip that person with an advanced measuring device that could image noninvasively the physiological activities of neurons inside someone else's brain. Given the opportunity to observe the electric impulses in someone else's brain during a concert, the born deaf person may be able to deduce that the other person under observation ``is experiencing the sounds of musical instruments,'' however, the born deaf person would not be able to imagine what hearing the sounds of the musical instruments would feel like. Thus, whatever meaning is attached by the born deaf man to the expression ``someone is experiencing sounds,'' it would be different and nothing like the meaning attached by a person with healthy hearing who had experienced sounds in the past \cite{Georgiev2017}.

\paragraph{What is it like to be a bat}
Consider bat's navigation in the natural world through sonar. Even though bats are mammals, hence closely related to us in the evolutionary history of life, no human is able to imagine what is it like to be a bat or what is it like to experience the surrounding world through reflected ultrasound waves. To say that from the physical recording of the electric activity of a bat's brain we have ``observed the bat's sonar experience'' could be accepted as true by definition but the actual, phenomenological, first-person bat's experience could have been subjectively very different, and yet, we still would have called it ``bat sonar experience.'' The expression ``bat sonar experience'' is well-defined with respect to whose experience it is, and regardless of what the actual bat experience is, or whether we can imagine what is it like to have it or not \cite{Nagel1974}.\\

One may try to redefine the term ``observation'' and insist that conscious experiences are observable in the sense that they are ``deducible'' from the observed data. For example, one may observe the physical electric impulses triggered in someone else's brain and deduce that the observed brain is experiencing pain, even though the observer himself is not experiencing pain. The ``deducibility'' of the pain, however, is not really an ``observation'' because it presupposes that the observer already has a kind of personal incommunicable knowledge of what is it like to feel pain, even before the deduction has been made. If the observer does not already know subjectively what is it like to experience pain, then the word ``pain'' would be either a label devoid of meaning or just a shorthand notation for the observed brain electric impulses. Redefinition of the term ``observation'' cannot show that experiences are observable, it can only corrupt the meaning of the word ``observation'' to ``utterance of word labels whose meaning we really do not understand.'' Thus, the inner privacy of conscious experiences is a problem that cannot be avoided and needs to be properly addressed by any physical theory. Insisting that ``the universe contains only observable entities'' is either false in the usual meaning of the term ``observation'' or meaningless if one attaches by definition the label ``observable'' to all existing things regardless of whether they can really be observed or not.

\begin{figure}
\begin{centering}
\includegraphics[width=160mm]{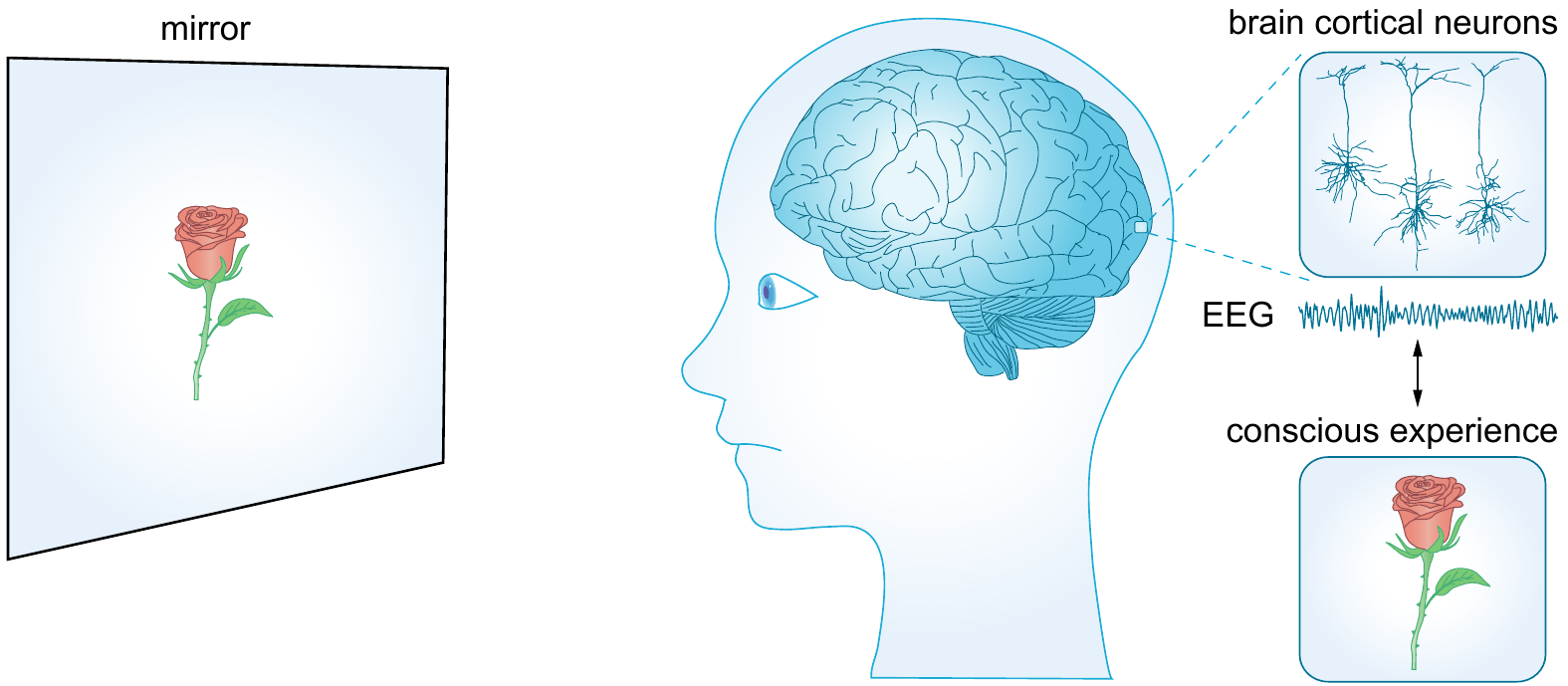}
\par\end{centering}
\caption{\label{fig:1}Illustration of an imaginary world in which conscious experiences happen to be observable from a third-person perspective. If we were experiencing the mental image of a red rose and we were able to take a look at ourselves in a mirror, we would see just the conscious experience of the red rose. In the real world, however, the conscious experiences are unobservable from a third-person perspective. If we consciously experience the mental image of a red rose and we are able to take a look at ourselves in a mirror, we will see the brain with its neurons and not the conscious experience. Because the brain looks quite different from what we experience, we are able to learn new things about ourselves with the use of a mirror.}
\end{figure}

\paragraph{An imaginary world with observable experiences}
In an imaginary world in which conscious experiences happen to be observable, looking at ourselves in a mirror would not have been remarkable for us since we already experience ourselves and know what we are experiencing (Figure~\ref{fig:1}). What would have been remarkable, however, is that we would have had direct access to the experiences of others by just taking a look at them. The paradoxical nature of such an imaginary world could be appreciated by contemplating upon the fact that such a world would have been just one global mind if it were a single connected network where everybody observes somebody and is observed by somebody \cite{Georgiev2017}.

\paragraph{The real world with unobservable experiences}
In the world we live in, the inner privacy of consciousness implies that if we happen to look at ourselves in a mirror, we will observe something different from what we are already experiencing. If a part of our skull is surgically removed for the sake of satisfying our own curiosity by taking a look in a mirror, we will see our brain reflected in the mirror, not our mind. Thus, by taking a look at ourselves in a mirror we are able to learn new things that we cannot learn solely from an introspective analysis of our own conscious experiences. For example, through accumulated observations of other brains, we have been able to discover that the brain is built up from neuronal networks with similar anatomical structure and organization across all individuals of the same animal species, providing evidence for inherited genetic information and common ancestry \cite{Darwin2006,Dawkins2004}.

\subsection{Incommunicability of conscious experiences}

\paragraph{Limits to the restoration of missing senses}
The inner privacy of consciousness is a direct manifestation of the fact that the subjective, first-person point of view of our conscious experiences is not communicable to others. If our conscious experiences were communicable to others through our communications, we would have been able to make a blind person see the world as we see it or a deaf person hear the world as we hear it. Unfortunately, we are unable to restore missing senses through communication alone. For example, we may communicate to a blind person that there is an obstacle in front of him, but this will not make the blind person see the obstacle. The blind person could understand the meaning of what an obstacle is due to the fact that already through another sense, such as touch, he has explored the world. The exploration of the world through touch allows one to form the abstract notion of location that does not depend on the exact sense through which a localized object is experienced. Thus, there may be some regularities within our conscious experiences that we could communicate to others. What we are unable to communicate to other people is the phenomenal nature of qualia, namely, what is it like to feel what we experience \cite{Georgiev2017}.

\paragraph{Locke's inverted qualia thought experiment}
Imagine a person who subjectively experiences yellowness when he is looking at a violet flower and blueness when he is looking at a marigold flower. The conscious experiences of that person would be qualitatively inverted compared to what you may experience when looking at the flowers violet or marigold \cite{Locke1850}. Because we neither have direct access to someone else's mind nor are we able to communicate to others what the blueness or the yellowness of our experiences are, we can never be sure that others do experience the same thing when put into an identical situation. Even worse, we can never be sure that others are able to experience what we are capable of experiencing in any situation. Thus, two people may both agree that they see a yellow marigold flower, even though one of them may have inverted qualia compared with the other. Moreover, each person is entitled to consider his quale to be normal because there is no objective test that can determine whose quale of yellowness is the normal one. Noteworthy, different qualia could be labeled with names e.g. ``yellowness'' or ``blueness,'' but these names (labels) do not carry any meaning (communicable knowledge) by themselves. The conscious subject perceiving the labels standing for different qualia fills in the first-person phenomenal perspective of what these qualia are and could be misled into wrongly believing that the labels themselves carry the qualia.\\

\paragraph{Impossibility to explain what the color experiences are to a color-blind person}
Color blindness results from mutations in the genes that produce retinal photopigments in the eye. Because color blind people do experience some color qualia when they look at certain colors, usually they are unaware of their condition before they get tested for color blindness in a medical check. What the medical tests can reveal is that the individual does not see a difference between two or more different colors that are distinguished by people who are not color blind \cite{Ishihara1972}. The tests cannot find, however, what color exactly is experienced by the color-blind person. If you were able to explain in words what color exactly you are seeing, you could have been able to cure color blind people only with your words. The empirical fact that you do not have the power to cure color-blind people by words implies that the phenomenal nature of conscious experiences is incommunicable.\\

The three philosophers from the Vienna Circle, Rudolf Carnap, Hans Hahn, and Otto Neurath, have succinctly summarized the scientific attitude toward the incommunicability of the phenomenal nature of qualia as follows:
\begin{quote}
``A scientific description can contain only the structure (form of order) of objects, not their `essence'. [$\ldots$] Subjectively experienced qualities -- redness, pleasure -- are as such only experiences, not [communicable] knowledge; physical optics admits only what is in principle understandable by a blind man too.'' \cite{Neurath1973}
\end{quote}

\section{Types of physical laws admitted by a private consciousness}

Confirmation that neither communication nor observation provides direct access to someone else's conscious experiences is obtained daily whenever you communicate with other people and observe them without actually experiencing what those other people experience. Even though conscious experiences are incommunicable and unobservable, there should be some physical laws in the physical theory of a subjective, private and unobservable consciousness, and these laws should be subject to constraints \cite{Georgiev2017}:

(1) There could be physical laws that postulate the existence of unobservable conscious minds. Thus, to each conscious mind could be attached a label such as $\Psi$ and questions related to that mind could be meaningfully asked. Different conscious minds and different conscious experiences could have different labels.

(2) There could be no psychophysical laws that explain what exactly is it like to feel the conscious experiences of a mind labeled as $\Psi$. Indeed, if it were possible to describe what the subjective, first-person point of view of conscious experiences is for the purposes of physical law, then since the physical laws in a theory are communicable, conscious experiences would have been communicable too.

(3) Because different conscious states within the physical theory cannot be theoretically differentiated by the incommunicable subjective nature of their content, they should be characterized and differentiated from one another by something communicable such as the probability distributions for different possible courses of action from which the conscious mind is able to choose. Because within deterministic theories the probability distributions are reduced to a single course of action, it follows that indeterministic theories have the capacity to differentiate between a larger number of possible conscious states $\Psi$ thereby allowing for a much richer physical theory of consciousness with the possibility of genuine free will.

Next, we will discuss the fundamental principles (axioms) that describe the physical reality in classical physics or quantum physics. We will show that the three constraints listed above on the possible physical laws of a theory of consciousness lead to paradoxes and logical inconsistencies in classical physics, but comply with all known quantum information theorems in quantum physics.

\section{Inner privacy of consciousness in classical physics}

The \emph{universe} is the totality of all existing things and \emph{physics} is the science that studies everything in existence. Because our minds exist in the universe, they should be described by physics and governed by physical laws \cite{Tegmark2015}. Unfortunately, even though physics is supposed to describe all existing things, current physical theories have nothing to say on whether an entity is conscious or not. Two historical reasons contribute to this state of affairs. First, philosophers of mind have persistently used the terms ``physical'' for the brain and ``nonphysical'' for the mind, without recognizing the paradox in defining physics as the natural science studying the universe and then branding all conscious minds in the universe as ``nonphysical'' as if they were outside the scope of physics. Second, physicists have habitually shied away from discussing consciousness and the difficult problems associated with it.

Identifying ``physics'' with classical physics is the main source of confusion in consciousness research. Classical physics refers to theories of physics that do not use quantization, including classical mechanics, classical electromagnetism and general relativity. The main principles (axioms) underlying classical theories are the \emph{observability} of all physical quantities, the \emph{communicability} of classical information at most at luminal speed, and \emph{determinism} governing the time evolution of physical states.

In classical physics, there is only classical information, which can be \emph{read}, \emph{copied}, \emph{multiplied}, \emph{broadcast}, \emph{recorded}, \emph{stored}, \emph{processed} with the use of irreversible Boolean gates and/or \emph{erased}. An example of classical information is the string of bits, 0s and 1s, encoding a digital movie recorded onto DVD. One can watch the DVD, copy the information from the DVD any number of times, and even erase the DVD in order to rewrite it with new information \cite{Georgiev2013}. The classical bits of information possess all of the above properties and naively these properties seem to be essential for our understanding of what information is. In Section~\ref{sec:inner-q}, however, we will show that quantum information does not have many of the properties of classical information and, most remarkable of all, quantum information is unobservable and incommunicable. But before we do that, let us see what kind of problems arise in a world of classical information in which everything is observable, communicable and deterministic.

Classically, the disturbing effect of any measurement (observation) could be made arbitrarily small. Hence, it is possible for an observation not to change the state of the physical system that is being observed and, due to determinism, it is possible to deduce with certainty what the physical state of the measured physical system was at the instant of time before the measurement was performed \cite{Susskind2013}. Also, all classical physical observables do have definite values at all times and these can be measured (observed) simultaneously. Thus, at least in principle, one can measure all physical observables of the classical brain and deduce the exact brain state at a given time. The knowledge of the current brain state could be then used to predict the brain state at any future moment of time.

\emph{Observability} of the state of every classical physical system poses a serious challenge to any classical theory of consciousness because the conscious experiences are unobservable. We are all aware that conscious experiences are only accessible from a first-person perspective and that we have no access to somebody else's experiences. The first-person accessibility of consciousness implies that we can never be sure whether any other animal species is conscious, or even whether another human being is conscious. Careful consideration of the principles underlying classical physics, however, leads to a paradox, namely, if everything that describes the state of a physical system is observable, how is it then possible that the conscious experiences are not observable? In other words, consciousness is banned from existence.

\emph{Communicability} of classical information poses yet another challenge to classical theories of consciousness. If everything about our conscious experiences were completely communicable, we would have been able to explain to others what is it like to experience the redness of a red rose or the smell of vanilla. Here, it is important to note that we do not question our ability to answer yes-or-no questions of the type: ``Are you consciously seeing a red rose?'' or ``Are you experiencing the smell of vanilla?'' What is questioned is our ability to communicate the phenomenal nature of qualia, namely, what is it like to have certain experiences from the first-person perspective.

Classically, if something is communicable then it is observable, and by contraposition, if it is unobservable then it is incommunicable. Ludwig Wittgenstein used two related arguments, namely, the private language argument and the beetle in the box argument, in order to highlight the problem. Later, Frank Jackson proposed the knowledge experiment and argued (inconsistently) that consciousness has to be ``nonphysical''.

\paragraph{Wittgenstein's private language argument}
Conscious experiences including my believing, seeing, imagining and loving are inner, private and inaccessible to anyone else. That very claim, however, is expressed in words that we all understand: ``believing,'' ``seeing,'' ``imagining'' and ``loving.'' We have learned these words from our parents or teachers through correcting incorrect uses and praising correct uses of the words \cite{Grim2008}. Thus, if consciousness is inaccessible to anyone else, then we should not have learned these words. Yet, we have learned these words; therefore it appears that consciousness has to be accessible \cite{Wittgenstein1999}.

\paragraph{Wittgenstein's beetle in the box argument}
Imagine that each of us had a beetle in a box into which no one else could look. If I say ``My beetle is fiddlededee,'' you may answer ``Mine is too'' or ``No, it is more flummadiddle than fiddlededee.'' Such conversation is nonsensical and the words like ``fiddlededee'' or ``flummadiddle'' can never acquire any meaning. Since mental terms possess meaning, it appears that consciousness cannot be private \cite{Wittgenstein1999}.\\

The private language argument and the beetle in the box argument show that there is a contradiction between the principles of classical physics and our introspective testimony of private consciousness. Unfortunately, rather than dismissing classical physics as incorrect, Wittgenstein predicted that there has to be a way for conscious experiences to become accessible to others. We have already shown, however, that if conscious experiences were accessible, then we would have been able to detect inverted qualia or make color blind people see colors through our communications. Since we do not have such abilities, we conclude that Wittgenstein's unconditional trust in the principles of classical physics is unfounded. We can associate our own experiences with the situations under which they are elicited, e.g. we can say ``yellow'' every time we see a marigold flower, but since our subjective experience of yellowness is inaccessible to others, those others could not possibly verify that their yellowness is the same quale as ours.

\paragraph{Jackson's knowledge argument}
Imagine a brilliant neuroscientist called Mary who has been raised and lived all her life in a black and white room where she was surrounded only by black and white objects. She has been well educated by reading all important books in neurophysiology and knows everything about vision, including how the incoming light of different wavelengths activates different red, green or blue cones in the retina, how this color information is converted into electric signals and delivered to the visual brain cortex where color is consciously experienced. Suppose that after having all the neurophysiological knowledge, Mary is allowed to see a red rose. By experiencing the redness of the red rose, did Mary learn something new? Intuitively it seems that by experiencing the first red object in her life, Mary acquires some kind of new knowledge that she did not have before. But how can this be if in the physical world there is only one kind of information, namely, classical information that can be perfectly well communicated and learned from books? Mary's room thought experiment was proposed in 1982 by Frank Jackson, and is now widely referred to as the knowledge argument, since it highlights the fact that there seem to be two kinds of knowledge, \emph{objective knowledge} that can be communicated to others and \emph{subjective knowledge} that cannot \cite{Jackson1982,Jackson1986}.\\

Classical approaches to consciousness have tried to address the knowledge argument by arguing that being in a certain mind/brain state is not the same as ``knowing'' what the mind/brain state is. Mary has to be in a brain state of experiencing the redness of the rose in order to experience her first red object. Before Mary experiences the redness of the red rose she knew in what brain state she had to be in, but she never was in that brain state while living in the black and white room. Such a statement is valid, but it completely misses the point of the knowledge argument, which is about the communicability of the knowledge of what is it like to experience something. After Mary experiences her first red object, she should form a memory trace in her brain that would inform her of what is it like to see red colored objects. If that memory of the red color were classical information, it would have been convertible into a string of bits, 0s and 1s, and could have been written in the books that Mary studied. Therefore, the main claim of the knowledge argument is that memories of phenomenal experiences could not be converted into communicable classical information. Otherwise, by simply sending strings of bits, 0s and 1s, we would have been capable of making people born blind know what seeing is and people born deaf know what hearing is.

\paragraph{Classical eliminativism is a self-refuting stance}
Accepting that classical physics is the correct description of the physical world implies that the information about all existing things is communicable. Consequently, one is forced to conclude that subjective, first-person conscious experiences are impossible. Indeed, the philosopher Daniel Dennett has argued that we are deluded about having conscious experiences and that we as conscious minds do not exist \cite{Dennett1991}. Entertaining as it may be, it would be irrational to believe the arguments produced by a non-existing mind or a mind deluded about its own existence. Furthermore, our memories of having had conscious experiences should not exist too because we do not exist, but then how are we able to remember things? Thus, if classical physics leads us to the conclusion that our consciousness is impossible, we should reject classical physics as inadequate. Mathematicians call this a \emph{proof by contradiction}---if you ever arrive at a paradoxical conclusion, reject the premise and do not trust the conclusion.

\paragraph{Classical functionalism leads to an evolutionary inexplicable consciousness}
Taking into account that anesthetized brains, unconscious brains or dead brains do not possess minds, some philosophers have proposed that consciousness is a product of the brain \cite{Bartlett2012,Fodor1981,Maudlin1989,Putnam1960,Putnam1975}. Unfortunately, the determinism of classical physics converts functionalism into epiphenomenalism. If the brain states produce conscious experiences, then these experiences cannot possibly have an effect upon brain dynamics, which is already fully determined by the physical quantities of the brain that enter into the fundamental equations of classical physics \cite{Georgiev2013}. Hence, consciousness is an epiphenomenon. For epiphenomenalists, our introspective feeling that our consciousness is causally effective is just an illusion produced by the brain \cite{Huxley1874}. However, according to the evolution theory something that is not causally effective cannot lead to evolutionary advantage and cannot be selected for by natural selection \cite{James1879}. In other words, classical functionalism renders conscious experiences utterly useless by reducing them to causally ineffective epiphenomena and asks us to believe that we are persistently deluded by our own introspective testimonies about what we are and what we can do \cite{Georgiev2019}.\\

Since classical physics leads to paradoxes in which either consciousness is banned or is rendered evolutionary inexplicable, it is rational to discard classical physics and study other empirically corroborated physical theories that supersede classical physics and support a novel kind of information, which is irreducible to classical information. Next, we will show how quantum physics and quantum information theory explain the inner privacy of our conscious experiences.

\section{Inner privacy of consciousness in quantum physics}
\label{sec:inner-q}

\subsection{States and observables in quantum physics}

At the beginning of the twentieth century, it became clear that classical physics fails to describe correctly the physical world, including but not limited to the spectral curve of blackbody radiation \cite{Planck1972}, the photoelectric effect \cite{Einstein1905}, stability of the hydrogen atom \cite{Bohr1981}, and diffraction of electrons in crystals \cite{Davisson1927,Thomson1927}. The concerted efforts of Nobel Laureate physicists, such as Max Planck, Albert Einstein, Niels Bohr, Louis de Broglie, Werner Heisenberg, Erwin Schr\"{o}dinger, Paul Dirac and Max Born, have demolished the foundations of classical physics through analysis of experimentally confirmed falsifications that have nothing to do with paradoxes in the theory of consciousness.

In 1926, Erwin Schr\"{o}dinger developed his wave equation \cite{OConnor2017}, which describes with astonishing precision the behavior of elementary physical particles
\begin{equation}
\imath \hbar \frac{\partial }{\partial t}|\Psi (\mathbf{r},t)\rangle =\hat{H}\,|\Psi (\mathbf{r},t)\rangle
\end{equation}
where $\imath$ is the imaginary unit, $\hbar$ is the reduced Planck constant, $\frac{\partial}{\partial t}$ indicates a partial derivative with respect to time, $|\Psi (\mathbf{r},t)\rangle$ is the wave function, $\mathbf{r}=\left( x,y,z \right)$ is the position vector, $t$ is time, and $\hat{H}$ is the Hamiltonian operator corresponding to the total energy of the quantum system \cite{Schrodinger1926,Schrodinger1928}.

The solution of the Schr\"{o}dinger equation is the quantum wave function $|\Psi (\mathbf{r},t)\rangle$ of the system, which is a continuous distribution in space and behaves like a vector in abstract Hilbert space as indicated by the vertical bar and angle bracket \cite{Dirac1967,Susskind2014}. At each point $\mathbf{r}$ in space at a time $t$, the value of the quantum wave function is a complex number $\Psi (\mathbf{r},t)$, referred to as a quantum probability amplitude. The absolute square $|\Psi (\mathbf{r},t){{|}^{2}}$ of each quantum probability amplitude gives a corresponding quantum probability for a physical event (e.g. detection of the particle) to occur at the given point in space and time \cite{Born1955}. If the state $|\Psi (\mathbf{r},t)\rangle$ of the quantum system is known, the quantum probabilities do not arise due to ignorance of what the state is but rather represent inherent \emph{propensities} of the quantum systems to produce certain outcomes under experimental measurement \cite{Popper1982}. Thus, the behavior of elementary physical particles is inherently indeterministic so that one cannot predict with certainty the future state of an individual particle, but only the probability with which a given future state could occur. In an indeterministic quantum world, epiphenomenalism is no longer unavoidable and the origin of the inner privacy of consciousness could be pinpointed in the physical properties of quantum information contained in the state vector $|\Psi (\mathbf{r},t)\rangle$ \cite{Georgiev2013}.

A characteristic property of the quantum wave function characterizing the quantum state vector  $|\Psi \rangle$ is that it is not observable. Contrary to the classical waves that can be observed in the ocean, the fabric of the quantum waves is woven from unobservable quantum probability amplitudes. What can be observed in a quantum measurement are the \emph{eigenvalues} of a measured physical observable $\hat{A}$, which is represented by a matrix as indicated by the hat \cite{Susskind2014}. Every observable $\hat{A}$  can act as an operator upon the quantum wave function and return (through matrix multiplication) another quantum wave function. For each observable $\hat{A}$, there is a set of proper wave functions called \emph{eigenvectors} such that if $\hat{A}$ operates on an eigenvector $|\Phi \rangle $ the outcome is the same eigenvector $|\Phi \rangle $ scaled by an eigenvalue $\lambda$, namely $\hat{A}|\Phi \rangle =\lambda |\Phi \rangle $. In general, a physical observable $\hat{A}$ of an $n$-level quantum system will have $n$ orthogonal eigenvectors $|{{\Phi }_{1}}\rangle ,|{{\Phi }_{2}}\rangle ,\ldots ,|{{\Phi }_{n}}\rangle $  with corresponding $n$ eigenvalues ${{\lambda }_{1}},{{\lambda }_{2}},\ldots ,{{\lambda }_{n}}$. Thus, the physical observable $\hat{A}$  can be spectrally decomposed
\begin{equation}
\hat{A}={{\lambda }_{1}}|{{\Phi }_{1}}\rangle \langle {{\Phi }_{1}}|+{{\lambda }_{2}}|{{\Phi }_{2}}\rangle \langle {{\Phi }_{2}}|+\ldots +{{\lambda }_{n}}|{{\Phi }_{n}}\rangle \langle {{\Phi }_{n}}|=\sum\limits_{n}{{{\lambda }_{n}}|{{\Phi }_{n}}\rangle \langle {{\Phi }_{n}}|}
\end{equation}

When the observable $\hat{A}$ is measured given a quantum system in an arbitrary state $|\Psi \rangle$, the only possible outcome of the measurement is a single eigenvalue from the set $\{{{\lambda }_{1}},{{\lambda }_{2}},\ldots ,{{\lambda }_{n}}\}$ and the quantum system collapses into the eigenvector corresponding to the obtained eigenvalue. The probability to obtain any given eigenvalue $\lambda _n$ is given by the \emph{Born rule} as the absolute square of the amplitude of the corresponding eigenvector $|{{\Phi }_{n}}\rangle $ projected onto the initial state $|\Psi \rangle$, namely
\begin{equation}
\text{Prob}({{\lambda }_{n}})=|\langle {{\Phi }_{n}}|\Psi \rangle {{|}^{2}}
\end{equation}

The unobservability of the quantum wave function originates in the fact that whenever we measure a physical observable $\hat{A}$, we also transform or prepare the system into an eigenvector $|{{\Phi }_{n}}\rangle $ of the measured observable. Hence, we learn what the quantum state of the measured system is after the measurement, but we cannot deduce with certainty what the quantum state $|\Psi \rangle$ was before the measurement \cite{Georgiev2017}. To see that, set the complex-valued quantum probability amplitudes resulting from the inner products as $\langle {{\Phi }_{n}}|\Psi \rangle ={{\alpha }_{n}}$, and apply the identity operator $\hat{I}=\sum\limits_{n}{|{{\Phi }_{n}}\rangle \langle {{\Phi }_{n}}|}$ to the initial quantum state before the measurement
\begin{equation}
|\Psi \rangle =\hat{I}|\Psi \rangle ={{\alpha }_{1}}|{{\Phi }_{1}}\rangle +{{\alpha }_{2}}|{{\Phi }_{2}}\rangle +\ldots +{{\alpha }_{n}}|{{\Phi }_{n}}\rangle =\sum\limits_{n}{{{\alpha }_{n}}|{{\Phi }_{n}}\rangle }
\end{equation}

If the quantum measurement generates the $n$-th outcome with probability $|{{\alpha }_{n}}{{|}^{2}}$, the quantum state is collapsed with absolute certainty of 1 to that corresponding eigenvector, namely
\begin{equation}
\sum\limits_{n}{{{\alpha }_{n}}|{{\Phi }_{n}}\rangle }\quad \to \quad 1|{{\Phi }_{n}}\rangle \label{eq:collapse}
\end{equation}

During the collapse process, all information about the other $n-1$ terms in the initial quantum superposition of $|\Psi \rangle$ is lost. It is the inherent propensity of quantum systems to generate indeterministic outcomes during measurement that effectively endows quantum systems with free will \cite{Conway2006,Conway2009,Georgiev2017}. In this sense, the free will manifested by quantum systems is intertwined with the unobservability of the quantum information encoded in the quantum state vectors \cite{Georgiev2017}.
In quantum information theory, the \emph{unobservability} of an unknown quantum state (i.e. state that is not prepared by us) could be proven rigorously as a theorem \cite{Busch1997}, and it could be further shown that unknown quantum states \emph{cannot be cloned} \cite{Wootters1982}, \emph{cannot be converted into classical information} \cite{Pathak2013}, \emph{cannot be broadcast} \cite{Barnum1996}, and \emph{cannot be deleted} \cite{Pati2000}.

\subsection{Consciousness as a quantum information-theoretic state}

\begin{figure}
\begin{centering}
\includegraphics[width=160mm]{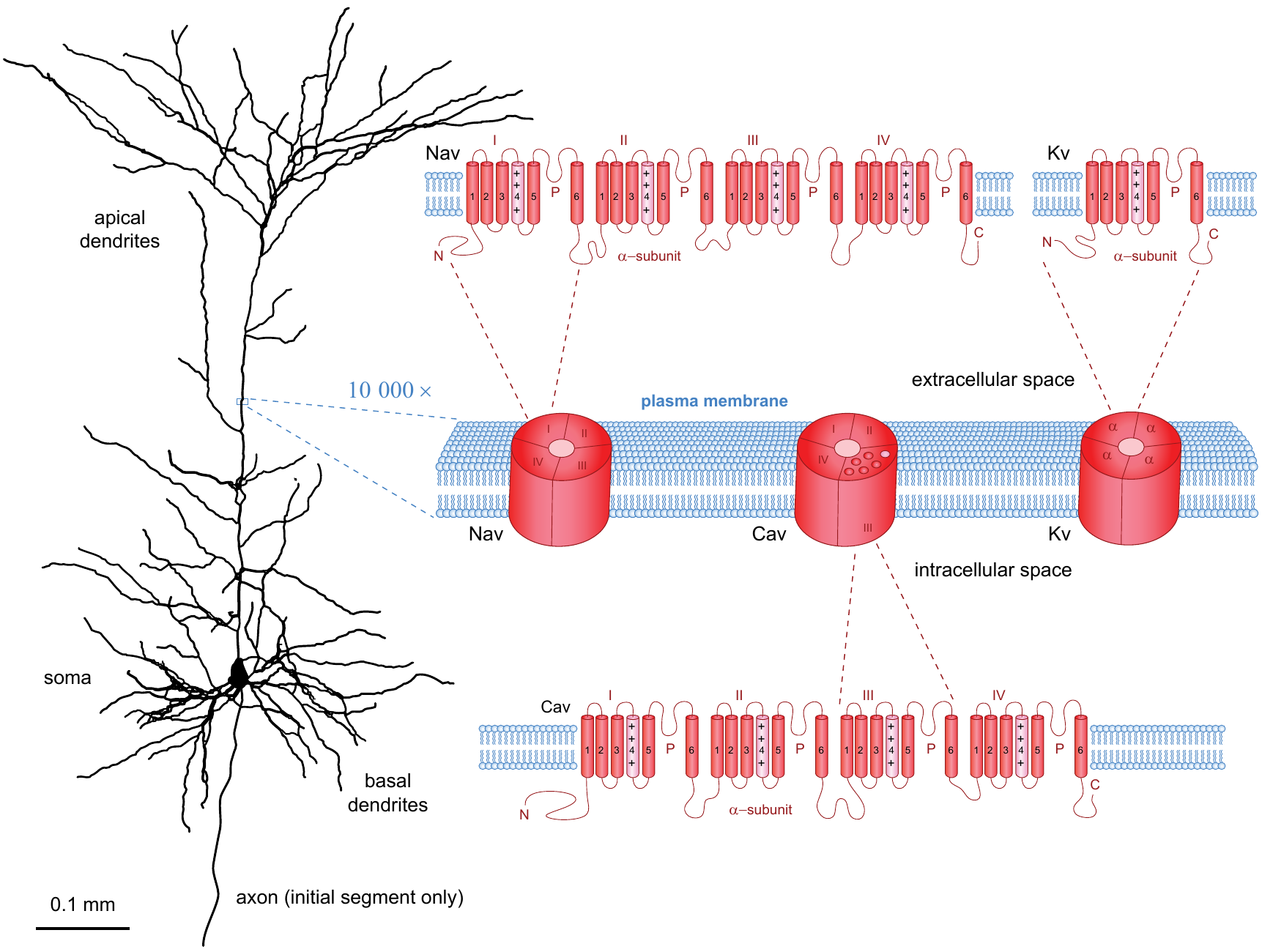}
\par\end{centering}
\caption{\label{fig:2}Morphology of a pyramidal neuron from layer 5 of rat motor cortex (NeuroMorpho.org NMO\_09565) and common structure of voltage-gated ion channels. Apical and basal dendrites receive synaptic inputs in the form of excitatory or inhibitory electric currents that summate spatially and temporally at the soma. If the transmembrane voltage at the axon initial segment reaches a certain threshold of depolarization around $-55$ mV, the neuron fires an action potential (spike) that propagates along the axon down to terminal axonal arborizations that form synapses mainly onto the dendrites (and rarely onto soma or axon) of target neurons. Neuronal electric properties are due to opening and closing of sodium (Nav), potassium (Kv) and calcium (Cav) voltage-gated ion channels. Structurally, each channel is built of four protein domains I--IV, each of which contains six transmembrane $\alpha$-helices (numbered from 1 to 6). The channel pore is formed by protein loops (P) located between the 5th and 6th $\alpha$-helices, whereas the voltage sensing is performed by the 4th electrically charged $\alpha$-helix within each domain. The scale bar of 0.1 millimeters applies only to the neuron reconstruction in the left panel. Since the thickness of the plasma membrane is only 10 nanometers, the right panel with the ion channels has an additional $10000\times$ magnification. Modified from \cite{GeorgievGlazebrook2014}.}
\end{figure}

Quantum physical states and the quantum physical observables are represented by two different kinds of mathematical objects in Hilbert space: the physical states are vectors $|\Psi \rangle$, whereas the physical observables are matrices (operators) $\hat{A}$. This fundamental difference, between what physically is and what can be physically observed, provides fertile ground in quantum physics for the accommodation of conscious experiences, which are subjective, private and inaccessible for external observers. Indeed, suppose that you are eating chocolate while undergoing open skull neurosurgery. What the surgeon would see is the pinkish-gray, walnut-shaped, jelly-like substance of your brain. If a microscope or other measuring devices are used, it would be possible for the surgeon to further zoom in on individual neurons (Figure~\ref{fig:2}) and record various complicated physical processes. But he would not be able to observe the taste of chocolate, because your experiences are inside your mind with a kind of insideness that is different from the way in which your brain is inside your head \cite{Nagel1987}. Namely, conscious experiences are unobservable. In classical physics, reductionism cannot work because the conscious mind cannot be identified with anything physical (e.g. the classical brain) as everything physical is observable. In quantum physics, however, identification of the conscious mind with the quantum information contained in the quantum state of the brain $|\Psi \rangle $ is possible, because $|\Psi \rangle$ is unobservable and fundamentally different from the observable brain $\hat{A}$ that can be measured, recorded, and examined with physical devices.
\pagebreak

Here, it should be noted that quantum nonlocality is able to provide a physical solution to the \emph{composition problem} (also referred to as the \emph{binding problem}) that afflicts classical mind--brain identity theories. Anatomically, the brain is composed of individual neurons cross-talking to each other. Endowing each neuron with a first-person conscious experience further requires an explanation of how conscious experiences of individual neurons are combined together in a single seamless stream of consciousness. Locality of classical physics leads to relativity of simultaneity and precludes instantaneous influences between spatially distant objects \cite{Bell1964,Sudbery2018}. The quantum law of composition \cite{Hayashi2015}, however, suggests a way out of the impasse.

The quantum state $|\Psi\rangle$ of a composite quantum system of $k$ components (with corresponding individual Hilbert spaces $\mathcal{H}_1$, $\mathcal{H}_2$, $\ldots$, $\mathcal{H}_k$) resides in a tensor product Hilbert space $\mathcal{H}=\mathcal{H}_1\otimes\mathcal{H}_2\otimes\ldots\otimes\mathcal{H}_k$. Factorizability of the Hilbert space does not imply factorizability of all quantum states. In fact, the vast majority of available quantum states $|\Psi\rangle$ in $\mathcal{H}$ do not admit factorization. Non-factorizable states for which there exists no decomposition in the form
\begin{equation}
|\Psi\rangle = |\psi_1\rangle\otimes|\psi_2\rangle\otimes\ldots\otimes|\psi_k\rangle
\end{equation}
are called quantum entangled states \cite{Ladyman2013}. An example of quantum entangled state of two spin-1 particles is
\begin{equation}
|\Psi\rangle= \frac{1}{\sqrt{3}}\left(|\uparrow_{z}\rangle_{A}|\uparrow_{z}\rangle_{B}+|0_{z}\rangle_{A}|0_{z}\rangle_{B}+|\downarrow_{z}\rangle_{A}|\downarrow_{z}\rangle_{B}\right) \label{eq:KS}
\end{equation}
where $|\uparrow_{z}\rangle$, $|0_{z}\rangle$ and $|\downarrow_{z}\rangle$ represent the eigenstates
of the $z$-component of the spin-1 observable $\hat{\sigma}$ with eigenvalues $\lambda = +1, 0, -1$, respectively, and indices $A$ and $B$ denote each of the two spin-1 particles. Applying the Kochen--Specker theorem to the state \eqref{eq:KS} for the components of the squared spin-1 observable $\hat{\sigma}^2$ along 33 different rays in the real 3-dimensional space, it can be shown that the measurement outcomes could not have been all pre-determined with absolute certainty (for a detailed mathematical proof we refer the reader to \cite{Kochen1967,Peres1991}). Moreover, if the two quantum particles $A$ and $B$ are separated a long distance away and are measured by experimenters with free will to choose the settings of their measurement devices, it can be concluded that the quantum particles possess free will too---the latter result due to Conway and Kochen is called the \emph{free will theorem} \cite{Conway2006,Conway2009}. Because the quantum entanglement leads to strong correlation between the outcomes of particles $A$ and $B$, it follows that the free will is manifested by the quantum entangled system as a whole, and not by its components individually. This means that conscious minds with a single integrated experience should be attributed only to pure non-factorizable quantum states in the brain \cite{Georgiev2017}. In other words, straightforward application of Conway--Kochen free will theorem to neuroscience supports quantum entanglement as the underlying physical mechanism that binds conscious minds together.

Identification of consciousness with quantum information comprising pure non-factorizable quantum states has important consequences for the evolutionary theory of human minds. If consciousness were a functional product of the brain, which is not identical to the brain, then consciousness would have been causally ineffective in the physical world as it does not enter into any physical equations, and ultimately human minds could not have been selected for their utility through natural selection. In order for natural selection to operate smoothly and produce complex minds, it has to operate on simpler ``mind stuff'' in the terminology of William James \cite{James1890}. Identity between consciousness and the quantum information of $|\Psi\rangle$ introduces the ``mind stuff'' directly into the Schr\"{o}dinger equation, which governs the dynamics of all physical systems, and makes consciousness causally effective since $|\Psi\rangle$ determines the probabilities of possible future courses of action (observable events) in quantum measurements performed by the environment. Thus, the quantum information-theoretic approach to consciousness endorses \emph{quantum panpsychism} according to which all quantum particles, including those in inanimate matter, are endowed with some form of elementary experiences. The importance of the human brain for sustaining human-like consciousness, however, needs to be highlighted in terms of memory and organization.

The importance of memory could be appreciated by considering the presence of conscious experiences in newborn babies, adults undergoing ketamine-induced dissociative anesthesia, or older people suffering from senile dementia.
Newborn babies are sentient but experience many of the sensory stimuli from the surrounding world for the first time due to lack of previously stored memories. Declarative memory and a sense of ``self'' develops gradually in the first 3 years of child's life, which provides an explicit example of how consciousness matures in incremental steps from elementary sentience \cite{Moscovitch1984}.
Dissociative anesthesia with ketamine provides yet another example of how conscious experiences may be present but the patient cannot remember them (amnesia). Many patients have their eyes open, exhibit spontaneous movements, and retain functional reflexes. Tears and saliva may also flow but the patient does not remember the operation or the anesthesia \cite{Trimmel2018}. The emergence from anesthesia itself may be associated with unpleasant dreams and hallucinations.
Older people with dementia further attest how the loss of past memories leads to gradual mood and personality changes \cite{Burns2009} that transform loved ones into poor, struggling with the disease, incapacitated, but nonetheless conscious creatures.

The importance of brain organization for sustaining human-like consciousness could be appreciated by considering dreams or drug-induced hallucinations. Rapid eye movement (REM) sleep is characterized with lucid, often illogical dreams, which may not be remembered at all unless we are awaken while dreaming. Because the electroencephalographic (EEG) activity during the REM sleep is suppressed ($\theta$ and $\beta$ waves) compared to wakefulness ($\gamma$ waves), it appears that disorganization of neuronal firing impairs memorization. Thus, the possible existence of non-memorizable, illogical, disorganized, chaotic, dream-like experiences in non-living matter postulated by quantum panpsychism does not imply animism, i.e. endowing inanimate objects with human-like consciousness. In addition, quantum information theory provides a natural explanation of why noxious drugs lead to illogical conscious experiences and hallucinations through disorganization of the physiological electric firing of brain neurons.

\subsection{Holevo's theorem constrains accessibility to the contents of quantum consciousness}

The inner privacy of consciousness is approached in two fundamentally different ways by functional theories of consciousness (functionalism) or identity theories of consciousness (reductionism). We will succinctly discuss the basic tenets of each of these two approaches.

Functionalism postulates that consciousness is a functional product generated by the brain, but which is not reducible to the brain. This means that consciousness does not appear in the physical equations that govern the dynamics of elementary brain constituents (particles). The essence of the functional approach is that consciousness emerges at a certain level of complexity of the brain dynamics and, as a ``product'', consciousness could have novel characteristics that are not possessed by the physical brain constituents. In such case, it may turn out that there is no fundamental prohibition on observing the inner world of consciousness at all. Also, the fact that we still have not learned to observe consciousness may turn out to be only a matter of technology and time. The drawback of functionalism, however, is that the physical laws and equations become completely useless for theoretical study of consciousness. Instead, since consciousness is not present in the physical equations, experimentalists should develop putative mind-reading technologies solely by trial-and-error and hope that their efforts will eventually pay off.

Reductionism postulates that consciousness can be identified with a physical entity inside the brain. This means that consciousness appears in the physical equations that govern the dynamics of elementary brain constituents. Consequently, consciousness is constrained by the physical laws and has the properties that can be deduced from general information-theoretic theorems. In such case, theoretical study of consciousness based on physical equations can be very useful to experimentalists for the design of their experiments and assessment of possible limitations faced by mind-reading technologies. The conundrum then is which set of physical laws, classical or quantum, should we rely on to describe consciousness.

Confronted with unobservable consciousness, philosophers have had a hard time explaining how in classical physics it is possible that we can communicate anything about our conscious experiences to others (Wittgenstein, 1999). If consciousness is composed of quantum information contained in quantum brain states \cite{GeorgievGlazebrook2018,GeorgievGlazebrook2019a,GeorgievGlazebrook2019b}, however, it would be possible to apply quantum information theorems to consciousness. Because quantum information cannot be fully converted into bits of classical information, the identity between consciousness and quantum information will imply that consciousness cannot be converted into classical bits of information, too. Indeed, qualia are not subject to exteriorization and we do not have a way to communicate in words or symbols what the phenomenal nature of different qualia is. Still, qualia can be introspectively compared and certain relationships between qualia can be encoded and communicated. For example, sounds can be loud or low, pleasant or unpleasant, etc., meaning that there is some order or regularity that can be captured in words and communicated, even though the phenomenal nature of each sound quale cannot be communicated. The classical understanding of information is very restrictive and incapable of reconciling the inner privacy of conscious experiences with the undeniable fact that we can talk about our experiences in a meaningful way (as exemplified by Wittgenstein's private language argument or beetle in the box argument). Quantum information theory, however, provides a deeper insight into the problem. Even though the quantum information carried by quantum systems is not observable and cannot be fully converted into classical information, each quantum system can carry a certain amount of accessible classical information subject to Holevo's theorem, namely a composite quantum system composed of $n$ two-level quantum subsystems (qubits) can carry up to $n$ bits of publicly accessible (extractable, measurable, observable) classical information \cite{Holevo1973}. Indeed, while non-orthogonal quantum states cannot be distinguished through measurement or observation, orthogonal quantum states can. A composite system of $n$ qubits resides in a Hilbert space with ${{2}^{n}}$  dimensions, which means that there are ${{2}^{n}}$ orthogonal states (basis states) potentially available for encoding of classical information. Specification of one basis state out of ${{2}^{n}}$ basis states delivers $H={{\log }_{2}}({{2}^{n}})=n$  bits of classical Shannon information \cite{Shannon1948}. Thus, something meaningful can be communicated in the form of classical information about conscious experiences that are generated by orthogonal quantum states. In essence, the quantum information is not completely inaccessible; there is some accessible part that is bounded by a certain amount of bits of classical information due to Holevo's theorem. Therefore, both the private language argument and the beetle in the box argument are deficient when viewed within the framework of quantum information theory. Only if we were able to say everything there is about our consciousness, it would have followed that consciousness is observable or accessible. From the fact that we can say something meaningful about our conscious experiences does not follow that we can say everything there is about these experiences. Hence, consciousness can be private insofar we can communicate only a limited number of meaningful things (classical bits) in regard to the content of our conscious experiences.

Demonstration that certain aspects of our conscious experiences are not communicable has already been provided by the inverted qualia thought experiment. In particular, we established that when we talk about our sensations, we never actually communicate the phenomenal aspect of the qualia associated with those sensations; rather we communicate the objective circumstances under which our sensations occurred. Further, it should be pointed out that while classical information is not firmly attached to its physical carrier (e.g. a poem can be carved in stone, written on paper, or encoded digitally in a computer file), quantum information is inseparable from its physical carrier (e.g. the quantum state of an electron cannot exist without a quantum physical electron). Consequently, the classical physical world cannot simulate the properties of quantum objects and classical computers cannot run quantum algorithms. This has important implications for the nature of human memories of past conscious experiences and the analysis of the knowledge argument. In particular, in order to be retrievable the memories have to be classical records of brain operations that lead to quantum brain states similar to those that have been associated with past experiences. If those classical records are communicated to or the operations are executed upon a brain that has not already had those experiences, the past experiences cannot be recalled or relived. Thus, quantum information theory provides a deep, mathematically precise explanation of why we cannot understand through communication what the bat sonar experience of the surrounding world is and why we cannot communicate to others the phenomenal nature of what we consciously experience.

Wittgenstein warned us that attempting to communicate the incommunicable is logically inconsistent:
``What we cannot speak about we must pass over in silence'' \cite{Wittgenstein1922}.
Asking how it is possible that incommunicable aspects of our conscious minds do exist, however, is a legitimate scientific question. The quantum information approach to consciousness explains the communicability/incommunicability of conscious experiences through the physical properties of brain quantum states, which are built up from quantum probability amplitudes. In doing so, it also provides a comprehensible analysis of certain philosophical problems (including the private language argument, the beetle in the box argument, and the knowledge argument) that appeared intractable within classical physics and vindicates the trustworthiness of our own introspective testimonies in regard to the nature of consciousness.

\section{Concluding remarks}

The fact that the quantum state is not observable and the inner world of consciousness is also not observable could point to an overlooked aspect of physical reality, rather than being just a coincidence. The reasons for unobservability appear to be different in both cases: the unobservability of the quantum physical state is a theoretical aspect of the mathematical model in Hilbert space, whereas the unobservability of conscious experiences of other minds is an empirical statement that we apprehend from our interaction with others.
To enforce logical consistency between the physical theory and empirical observations, however, we have explored the implications of a possible reductive identification between consciousness and the quantum information contained in pure non-factorizable quantum states. While this quantum approach may turn out to be incomplete and wanting, similarly to older classical approaches, the main thesis of this work is that it provides a fresh perspective on the mind--brain problem and equips theoreticians with a toolset of quantum information-theoretic no-go theorems that enrich the number of conceivable solutions to the problem.

In classical physics, everything is observable and communicable. Consequently, there is either no room left for unobservable and uncommunicable consciousness, or if such consciousness is postulated to emerge as a functional product of the brain, then due to the determinism of classical physics the emergent consciousness will be a causally ineffective epiphenomenon. In an indeterministic quantum world built up from unobservable and incommunicable quantum information, however, epiphenomenalism is no longer unavoidable and the origin of the inner privacy of consciousness could be pinpointed in the unobservable and incommunicable quantum information contained by the quantum state of the brain \cite{Georgiev2017}. The observable brain then is the accessible classical information about the brain that can be extracted by physical measuring devices bounded by Holevo's theorem.
Thus, we have shown that the conscious mind is not ``nonphysical'' and modern advances in quantum information theory are directly applicable for the analysis of classically unresolved thought experiments that led to paradoxes of consciousness. In essence, quantum mechanics plays a non trivial role in human thinking and consciousness, which goes beyond the deficiencies of classical physics \cite{Matsuno2000,Matsuno2006,Melkikh2015,Melkikh2019}.

It is often the case that foundational research affects a wide range of disciplines without immediately obvious practical applications that benefit humankind. The inner privacy of consciousness, however, is a subject of great interest and growing concern in a digital world where our daily lives are tracked through internet searches, public camera recordings or triangulation of mobile phone signals and Wi-Fi devices. Selling our private data by internet-related corporations to third parties is a lucrative business that profits from exploiting our personal biases when purchasing services and spending our funds. The promise of newly developed technologies for mind-reading adds only fear that societies in the near future will lack any privacy whatsoever \cite{Eggers2013} and individuals will lose completely their freedom due to increased predictability of their actions. Here, we have shown that fortunately for us, the fundamental quantum physical laws set bounds on what can be ultimately observed and extracted in the form of classical information from measurements performed upon our quantum brains. While the physical boundary of privacy guaranteed by quantum physical laws is narrowed to the extent of our brains and certainly does not solve the problem of digital or other kinds of privacy in our daily lives, it provides some comfort in knowing that individuals should not succumb to external blackmailing if threatened with alleged mind-reading devices \cite{Elga2004,Georgiev2013}. By raising the awareness of consciousness researchers for the utility and importance of quantum information theory, we hope to stimulate further interdisciplinary research and discussion on the physical relationship between the conscious mind and the observable brain.

\section*{Acknowledgements}

I would like to thank two anonymous reviewers for their critical comments, which have helped to improve the clarity and presentation of the main ideas.

\section*{Conflict of interest}

The author declares that he has no conflict of interest.

\end{document}